# An Explainable AI Model for Binary LJ Fluids

Israrul H Hashmi[1], Rahul Karmakar[1,2], Marripelli Maniteja[1], Kumar Ayush[1] and Tarak K. Patra[1,2*]

[1]Department of Chemical Engineering, Indian Institute of Technology Madras Chennai, TN 600036, India

[2]Center for Atomistic Modeling and Materials Design, Indian Institute of Technology Madras Chennai, TN 600036, India

## Abstract

Lennard-Jones (LJ) fluids serve as an important theoretical framework for understanding molecular interactions. Binary LJ fluids, where two distinct species of particles interact based on the LJ potential, exhibit rich phase behavior and provide valuable insights of complex fluid mixtures. Here we report the construction and utility of an artificial intelligence (AI) model for binary LJ fluids, focusing on their effectiveness in predicting radial distribution functions (RDFs) across a range of conditions. The RDFs of a binary mixture with varying compositions and temperatures are collected from molecular dynamics (MD) simulations to establish and validate the AI model. In this AI pipeline, RDFs are discretized in order to reduce the output dimension of the model. This, in turn, improves the efficacy, and reduce the complexity of an AI RDF model. The model is shown to predict RDFs for many unknown mixtures very accurately, especially outside the training temperature range. Our analysis suggests that the particle size ratio has a higher order impact on the microstructure of a binary mixture. We also highlight the areas where the fidelity of the AI model is low when encountering new regimes with different underlying physics.

# I. Introduction

The LJ potential is widely used to model the interaction between atoms or molecules of a wide range of materials. The classical LJ potential describes the interaction between a pair of particles as a combination of repulsive and attractive forces.[1–4] This simple yet effective model has been instrumental in studying a variety of physical systems, particularly in the fields of molecular dynamics and statistical mechanics. Binary LJ fluids, consisting of two different particle types, pose an additional layer of complexity due to the interplay between the interactions within each species and between species. These fluids can model real-world systems such as liquid mixtures, binary gases, or alloys. The phase behaviour of binary systems are intricately connected to their compositions viz., mole fraction, size ratio, and cross-interaction along with thermodynamic and environmental conditions.[5–16] Traditionally, many liquid state theories are used to calculate the radial distribution of atoms in a material system.[17–21] Subsequently, molecular simulations have been progressively used to determine radial distribution function of molecule systems.[22–24] Accurately computing all possible distributions of particles, estimating phase behavior, critical points and other thermodynamic properties for a wide range of conditions and environment using molecular simulations could be computationally expensive and challenging also, especially at low temperature region, wherein it is quite challenging to achieve thermodynamic equilibrium. Therefore, we aim to examine the capabilities and limitations of AI to explore the composition space of a binary LJ fluid for a range of thermodynamic conditions.

The AI has shown tremendous potential in studying complex fluids and soft matter systems.[25–30] These models can not only interpolate within known data but seem to extrapolate and predict behaviors and properties outside the observed parameter space, making them promising tools in situations where experimental or simulation data are limited.[31,32] In this work, we explore the application of AI models to binary LJ fluids, focusing on key methodologies and out of range prediction. We focus on predicting the RDF of a binary fluid outside the thermodynamic conditions of the training data. We have recently established an AI pipeline for predicting the RDF of multicomponent systems as a function of their compositions.[33,34] These models predict the RDF of a pair of particles in a multicomponent system for a given set of the size ratio of the two types particles, the concentration of one-type of particle, and the interaction between the two types. The output dimension of these models are high as it predicts a function rather than a single value. Typically, the fidelity of ML models tends to decrease as the output dimension increases due to several challenges. As the number of output variables grows, the model requires significantly more data to learn meaningful patterns. With limited data, the model struggles to generalize well. High-dimensional outputs often lead to more complex models that may overfit the training data, reducing their ability to generalize to unseen data. Training a model with a large output space increases computational complexity and may make the optimization harder, leading to suboptimal convergence. Also, in high-dimensional spaces, the loss function may become less informative, making it difficult for the model to distinguish between good and bad predictions effectively. Secondly, it is not well known how effectively the AI RDF prediction model can extrapolate. Furthermore, the explainability of RDF prediction models remains limited. In this study, we aim to address these challenges by pursuing the following three interconnected objectives: (1) developing an efficient RDF prediction model, (2) evaluating its extrapolation capability, and (3) gaining a deeper understanding of composition-property correlations of the binary LJ fluid via AI modelling.

To address the above questions, we perform MD simulations of the binary LJ fluid for a varied range of composition and temperatures. We use these data to build an AI model viz., deep neural network (DNN) that can predict RDFs of the fluid mixture. Within this AI framework, we discretize an RDF into multiple points. It serves two purposes. First, it reduces the output dimension of the DNN to one. Second, it increases the number of data points. Therefore, a problem originally formulated as a mapping from an X-dimensional input vector to a Y-dimensional output vector is transformed into a mapping from an (X+1)-dimensional input space to a single scalar output. This reformulation simplifies the learning process by reducing the complexity of multi-output prediction, enabling models to focus on capturing underlying relationships with enhanced accuracy and generalizability. Since, the input and output dimensions are low, the overall efficiency and accuracy of the model is significantly better than previous RDF prediction models. We also include the thermodynamic condition viz., temperature as an input to the model, which was missing in previous approaches. Hence, the input dimension of the present model is seven, corresponding to the size ratio of the two types of particles, the concentration of one type of particle, their cross-interaction, temperature, interparticle distance, and two labels representing the types of particles. The output dimension is one, representing the RDF value corresponding to the input vector. We find the present AI model is better suited for limited data as well as capable of extrapolation. We further perform SHAP[35] analysis to interpret the predictions of the model. The SHAP is a technique for explaining individual predictions made by an AI model by attributing the prediction to contributions from each input feature. For each instance in the dataset, the contribution of each feature to the prediction varies depending on the specific values of the features in that instance. It appears that the size ratio of the two type of particles of a binary fluid has the highest impact on determining the spatial distribution of particles in the system. We compliment the SHAP analysis by lower dimensional project of RDFs and cluster analysis. We also investigate thermophysical scenarios in which models encounter difficulties in accurately predicting the properties of the LJ binary mixture. These challenges arise due to the complex intermolecular interactions, diverse compositional variations, and the nonlinear nature of thermodynamic properties within such systems.

## II. System and Data Generation

We conduct MD simulations of an LJ binary fluid mixture, whose constituents are named as A and B-type particles. In this model system, the pair interaction between two particles is considered as $V(r) = 4\epsilon_{ij}\left[\left(\frac{\sigma_{ij}}{r}\right)^{12} - \left(\frac{\sigma_{ij}}{r}\right)^6\right]$. Here the $i$ and $j$ refer to the type of a particle – A or B. The $\epsilon_{ij}$ and $\sigma_{ij}$ are the interaction strength and effective size of two interacting particles. The LJ interaction between a pair of particles is truncated and shifted to zero at a cut-off distance $r_c = 2.5\sigma_{ij}$. There are three pairs of interaction in the system. We keep $\epsilon_{AA} = \epsilon_{BB} = \epsilon$, the unit of energy. The cross interaction energy $\epsilon_{AB}$ is varied from $0.2\epsilon$ to $1.\epsilon$. We consider particle size $\sigma_{AA} = \sigma$, which is the unit of length in our calculations. The $\sigma_{BB}$ is varied from $0.5\sigma$ to $2\sigma$. Thus, the A and B type particles size ratio $S = \sigma_{BB}/\sigma_{AA}$ is varied from 0.5 to 2.0. The length scale for the cross interaction potential is chosen as $\sigma_{AB} = (\sigma_A + \sigma_B)/2$. We simulate the system for various compositions of A and B type particles. The system consists of 10000 particles. The fraction of B-type particles ($C_B$) is varied from 0.01 to 0.5. The MD simulations are performed in an isothermal isobaric ensemble (NPT), wherein the temperature and pressure are controlled by the Noose-Hoover thermostat and barostat, respectively. We use the velocity-Verlet algorithm with a timestep of $0.005\tau$ to integrate the equation of motions. Here, $\tau = \sqrt{m\sigma^2/\epsilon}$ is the unit of time; $m$, $\sigma$ and $\epsilon$ are the unit of mass, size and interaction energy, respectively. Simulations are conducted at a reduced temperature $T^* = k_B T/\epsilon$ where $T$ and $k_B$ are the temperature and the Boltzmann constant, respectively. We vary T* in a range of 0.6 to 1.4. The reduced pressure $P^* = P\sigma^3/\epsilon = 1$ is kept constant for all the simulations. All the systems are equilibrated for $10^7$ steps and followed by $10^7$ steps of production runs.

. All the simulations are conducted within the MD environment of LAMMPS[36] package. We calculate RDFs using 1000 configurations from the production runs. All the RDFs are up to an interparticle distance of $10\sigma$. The training data are collected over the temperature range from $T^* = 0.8$ to $T^* = 1.2$. The test data are collected from the MD simulations for $T^* = 0.6$ and $T^* = 1.4$. We compute 900 RDF for training and 600 RDFs for testing the model's performance. An MD snapshot and the corresponding RDFs of the system are shown in Figure 1A and 1B, respectively. Each RDF curve corresponds to 100 discrete points. Therefore, we have 90000 data points for training the model and the test data size is 60000.

## III. Construction of the DNN Model

The fingerprint vector, which is the input to the model, is of size seven. They are schematically shown in Figure 1D. The first four entries in the fingerprint are temperature, $T^*$, the fraction of B-type particles in the system, $C_B$, particle size ratio, S, and A-B interaction, $\epsilon_{AB}$. The subsequent two entries define the target pair type for the RDF. A value of 11, 10 and 01 correspond to AB, AA and BB pairs, respectively. Lastly, the interparticle distance (*r*) between a pair of particles is considered as the 7[th] entry in the fingerprint vector. As the size of the fingerprint vector is seven, the input layer of the DNN consists of 7 nodes, each representing one entry of the fingerprint vector. The output layer of the DNN consists of one node. It represents the *g(r)* value for a given interparticle distance *r*. The DNN has 5 hidden layers. The number of hidden layers and their sizes are decided based on initial trials in order to optimize the performance of the model. We found 7-112-56-28-14-7-1 as the optimal DNN for this binary fluid RDF problem. We use the rectified linear unit (ReLU) function[37] as the activation function of the DNN model. A standard backpropagation algorithm is used for the network training. A gradient-based stochastic optimization algorithm viz. the Adam optimizer, is used to optimize the parameters of the DNN during the backpropagation.[38] The DNN model is built within the Keras API environment.[39] We use 0.001 as the learning rate during the training of the model. The loss function, which is defined as the mean square error in the output with respect to the ground truth, is shown in Figure 2A during the training of the model. It shows how the DNN prediction improves over training cycles. We run ~1000 training cycles during which the loss function reaches a plateau.

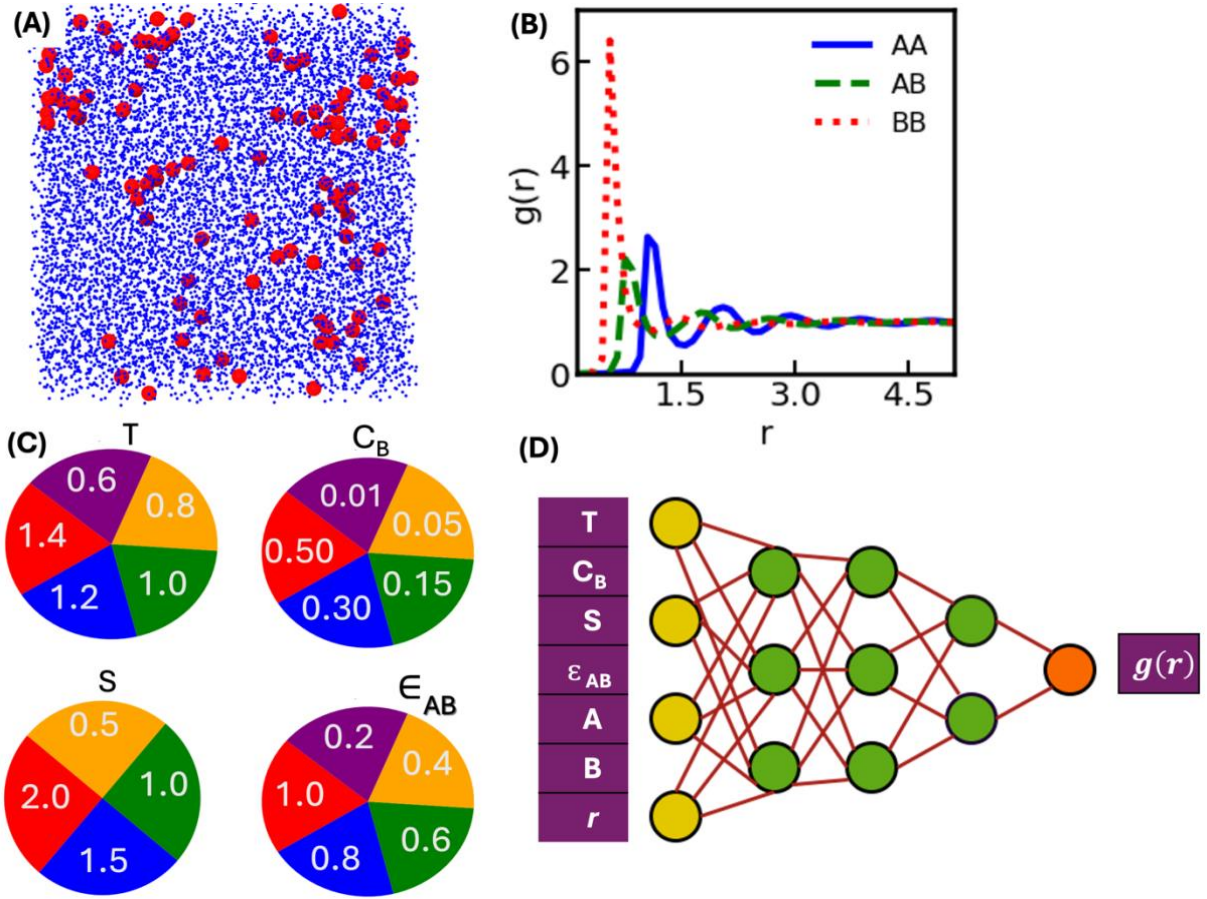

*Figure 1: An MD snapshot of the binary fluid is shown in (A). Three possible radial distribution functions of the system is shown in (B). The composition space of the system is schematically shown in (C), wherein T, $C_B$, S and $\epsilon_{AB}$ are the temperature, B-particle concentration, the size ratio of the two types of particles and the A-B interaction strength, respectively. The fingerprint vector and the DNN model are schematically shown in (D).*

## IV. Results and Discussion

We randomly split all the data into training and test sets. The DNN is built with the training set. The loss function during the training process is shown in the Figure 2A. The coefficient of determination $R^2$ is 0.99 for training and test data sets. This model outperforms our previous

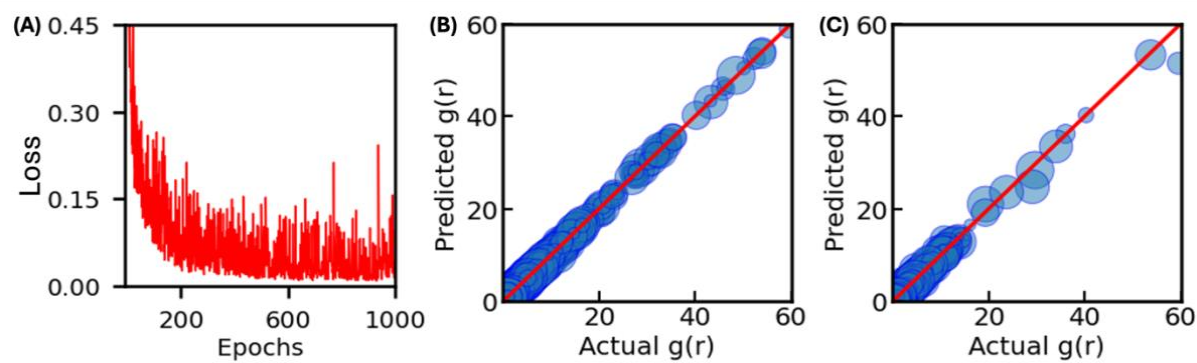

*Figure 2: Interpolative DNN model. The loss is shown as a function of the training cycle in (A). The predicted g(r) values are plotted against their actual values in (B) and (C) for the training and test data sets, respectively. The $R^2$ is 0.99 for both the cases.*

RDF prediction model.[33] The current RDF model takes about ~40 seconds of CPU time for training 100 RDFs. This is significantly faster than the previous RDF model[33], which takes ~

250 seconds of CPU time to learn 100 RDFs. The discretization of RDFs and the learning and prediction of individual RDF values, rather than the entire function, help reduce the model's complexity and improve efficiency. Next, we build another model for the extrapolation task. This extrapolative model is trained with the RDF data that are collected over the temperature range of $T^* = 0.8$ to $T^* = 1.2$. The loss function of this model during its training is shown in Figure 3A. We test the model's performance for two outside range temperatures viz., $T^* = 0.6$ and $T^* = 1.4$. This time, the model achieves a coefficient of determination $R^2 \sim 0.98$ and $R^2 \sim 0.89$ for the training and test data sets, respectively. The training and test parity plots are shown in Figures 3B and C, respectively. Further, we show the performance of the model for different types of g(r) for the test data set in Figure *3D-F*. We observe that the model predicts AA and AB types of g(r) functions with significantly high accuracy $R^2 \sim 0.97$ and $R^2 \sim 0.98$, respectively. This suggests our model is able to predict out of range temperature data - both high and low ends with reasonably high accuracy. However, the model performance for the case of BB type RDFs is relatively poor. The $R^2$ for the BB pair RDFs is $\sim 0.83$. We observe a few predicted BB pair RDF values deviate significantly from their actual value. This reduces the overall model performance for BB pair RDFs. To better understand this discrepancy, the predicted RDFs as a function of the interparticle distance is compared with the corresponding ground truth in Figure 4 for a few representative cases. It suggests that the model has largely captures the long range pair correlations for all the pairs of the binary mixture with an exception of the first peak in BB of Figure 4c, which corresponds to the lowest temperature. We also chose another composition for this extreme temperature. We plot the actual and predicted RDF for $T^* = 0.6$, $C_B = 0.01, S = 1.5, \epsilon_{AB} = 0.2$ in Figure 4A. In this case, the model is able to identify the 1st peak position; however it could not quite capture the 2nd and 3rd pack of the RDF. In addition, the 1st peak height is significantly underpredicted. Figure 4B represents the MD snapshot of the system for this specific composition and temperature for visual inspection. It clearly shows that the *B*-type particles are forming small clusters, which is also evident from the large 1st peak height of the actual BB pair RDF. We also observe that at the nearest temperature, $T^* = 0.8$, B particles are reasonably mixed well with the A particles. Therefore, the system undergoes a phase transition (mixing to demixing) as the temperature decreases from 0.8 to 0.6. (Figure 4). These structures are absent in the training data. In such situation, the model struggle to predict the RDF. This suggests that the model learns specific patterns in the training data very well, making them less effective when faced with new physics. It affirms that the model's ability to extrapolate is contingent on the continuity of governing physical principles. However, extrapolation in AI and computational modeling is often unreliable, especially when encountering new regimes with different underlying physics. Thus, the occurrence of a phase transition at low temperature in the present system reduces the fidelity of the AI model.

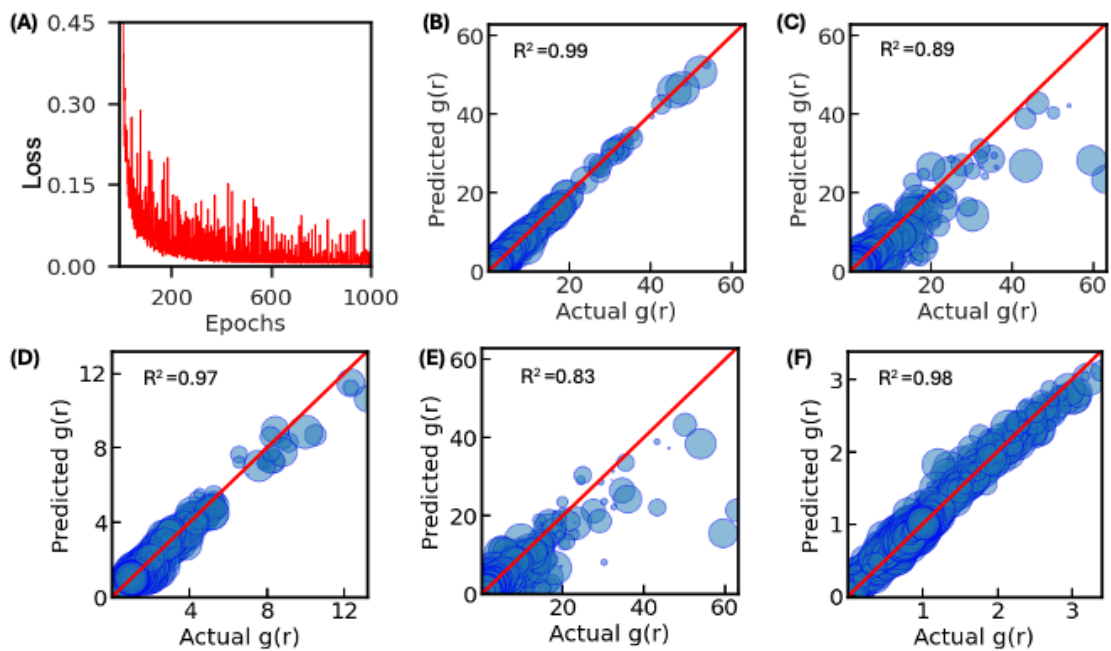

*Figure 3: Extrapolative DNN model. The loss function of the model during the training is shown in (A). The predicted g(r) values are plotted against their actual value in (B) and (C) for training and test data sets, respectively. Individual parity plots for AA, BB and AB pairs g(r) values are shown in (D), (E), and (F), respectively.*

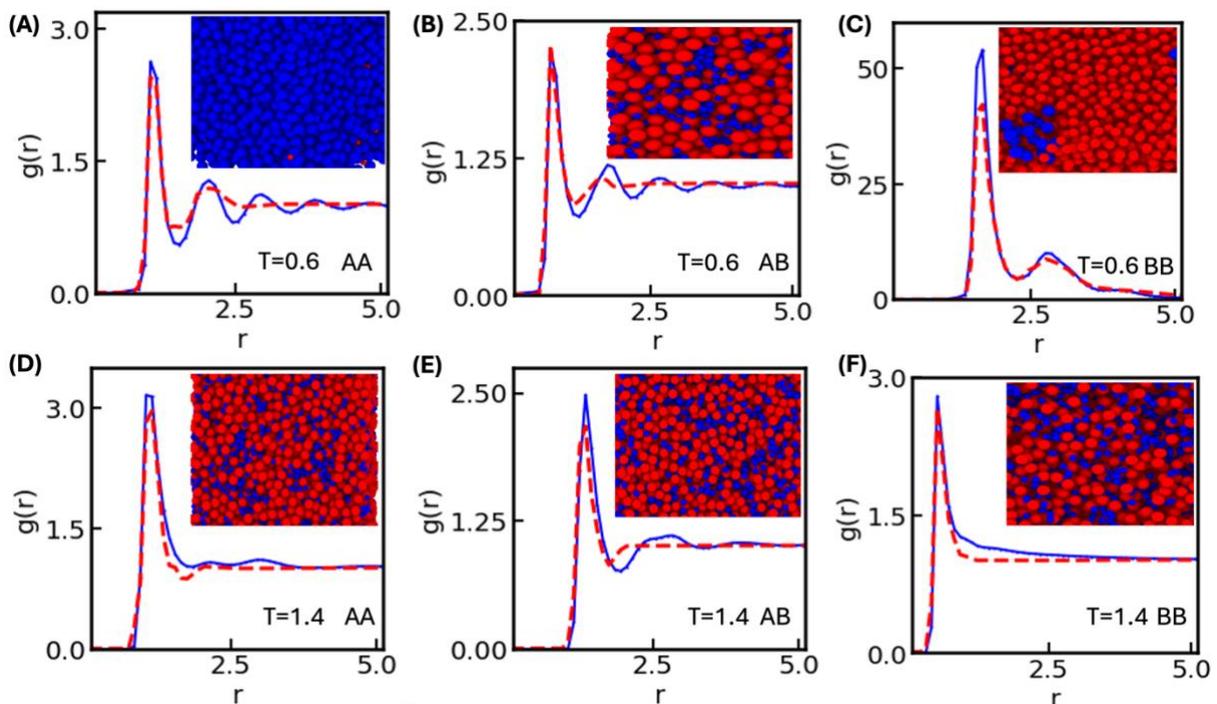

*Figure 4: Actual (blue solid line)l and predicted (red dotted line) AA pair g(r) functions are shown in (A) and (D) for $T^* = 0.6$ and 1.4, respectively. Similarly, the AB pair g(r) functions are shown in (B,E), and the BB pair g(r) functions are shown in (C,F). The MDs snapshots of corresponding systems are provided in the inset of all the panels for visual inspections.*

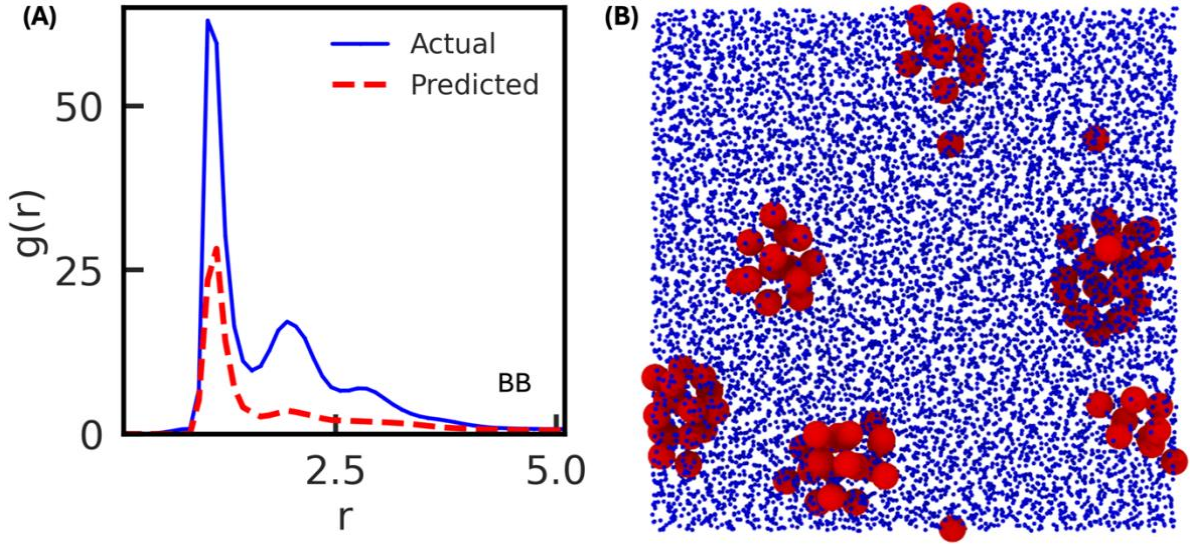

*Figure 5: Actual and predicted g(r) functions for the BB pair at $T^* = 0.6, C_B = 1, S = 1.5, \epsilon_{AB} = 0.2$ are plotted in (A). The corresponding MD snapshot is given in (B), where red beads are B-type and blue beads are A-type particles.*

Next, we study the extrapolation capability of the model as a function of training data points. We randomly pick a subset of data from the training set, and build several models. We use these models to predict the g(r) values of the test data set. Figure 6A shows the $R^2$ as a function of training data points. The $R^2$ increases rapidly with the volume of training data. The rate of increment slows down at the higher limit of training data. The predicted BB pair RDF function is compared with the actual BB pair RDF function for two representative cases in Figure 6B and C. They correspond to 50000 and 90000 training data points, respectively. The pair correlations is mostly missing in model's prediction for the former case. It captures the correlation significantly for the latter case. Overall, the $R^2$ is about 0.1 for the case of 50000 training data points, while it is about 0.9 for 90000 training data points. Thus, the fidelity of the model for out of range prediction improves with training data.

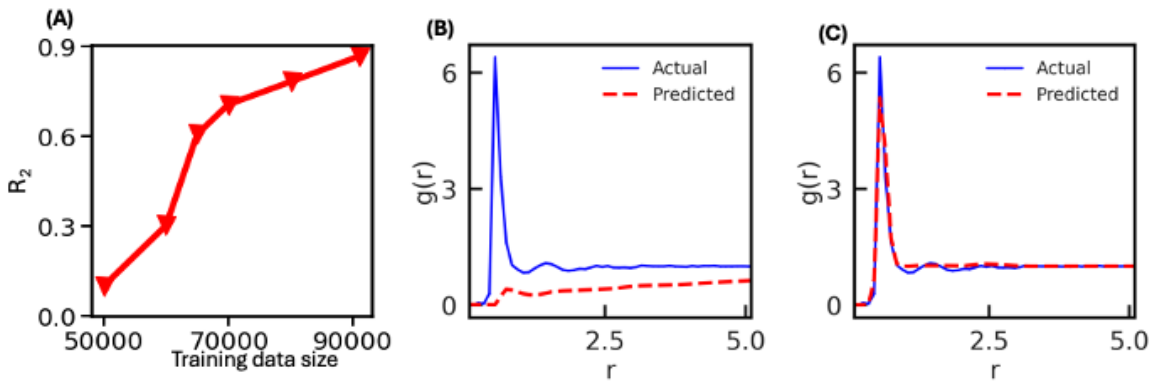

*Figure 6: The $R^2$ of the model for the test data set is plotted as a function of the training data size in (A). The precited and actual g(r) functions are compared in (B) and (C) for smaller training data set, and large training data set, respectively.*

Now, we focus on the explainability of the model. We use uniform manifold approximation and projection (UMAP)[40] to map all the RDFs in a two-dimensional (2D) space in Figure 7. We intend to understand which feature—temperature, particle size ratio, concentration, or cross-interaction—plays the most prominent role in determining the particle distribution in the mixture. The UMAP is a nonlinear dimensionality reduction technique that enables the visual inspection of high-dimensional data in a 2D representation. This lower-dimensional representation captures salient features in the dataset and provides a better understanding of the data distribution and any systematic patterns that may be present in the data. We concatenate three g(r) functions, which are for AA, AB, BB pairs, in one array for a particular combination of $T^*, C_B, S$ and $\epsilon_{AB}$. This array serves as the input to the UMAP. Each g(r) starts from $0.9\sigma$ and at $5\sigma$. This ensures that the number of null values in the array is minimized and that it captures the region where the function varies, beyond which it remains mostly unchanged. The dimension of the single array is 240, which describes the whole system completely for a particular combination of $T^*, C_B, S,$ and $\epsilon_{AB}$. The UMAP of all the RDFs in 2D are shown in Figure 7A. Further, we perform K-mean clustering of data in the 2D space. We observe four optimal clusters in this space based on the inertia and Silhouette scores. The inertia and Silhouette scores are plotted as a function cluster size in Figure 7B. The inertia decreases with increasing cluster size. The optimal cluster size is defined as the elbow point where the decrease in inertia begins to slow in Figure 7A, this corresponds to K=4. Similarly, the Silhouette measures how each data points fit into a cluster and how distinct it is from other clusters. The highest score of the Silhouette curve is considered as the optimal cluster size, which is K=4 as shown in Figure 7B. Here, both parameters – inertia and Silhouette suggest four distinct clusters of points in the UMAP. We assign a distinct color to each of the four clusters in Figure 7A. Now, we label all the points in the UMAP with their respective features values ($T^*, C_B, S, \epsilon_{AB}$), and calculate the probably of a particular feature being present in a cluster ($p_i$). We then compute the mixing entropy of a feature as $E = -\sum_i p_i \log(p_i)$, where $i$ represents all possible values of a feature. We calculate average E over all the clusters for all the features. Here, the mixing entropy of a feature quantifies the extent of mixing of all its values in a cluster. We observe average entropy is lowest for S and highest for $T^*$ (Figure 7C). It suggests $S$ is the most important feature to get distinct particle distributions of the system. In other words, the size difference between the two type of particles plays most dominant role in determining the particle distribution in the system. We further perform the SHAP[35] analysis of the DNN model to rank all the four features in terms of their importance in determining RDFs of the binary mixture as shown in Figure 7D. The SHAP score for the particle size ratio, $S$, is found to be highest among all the features, followed by the cross interaction $\epsilon_{AB}$. This supports the fact that the particle size ratio is most dominant mode to achieve a wide range of particle distribution in a binary LJ fluid. Moreover, these AI analysis are in general agreement with recent experiments and theories that indicate the size ratio has the strongest effect on the packing of particles in binary mixtures.[41,42]

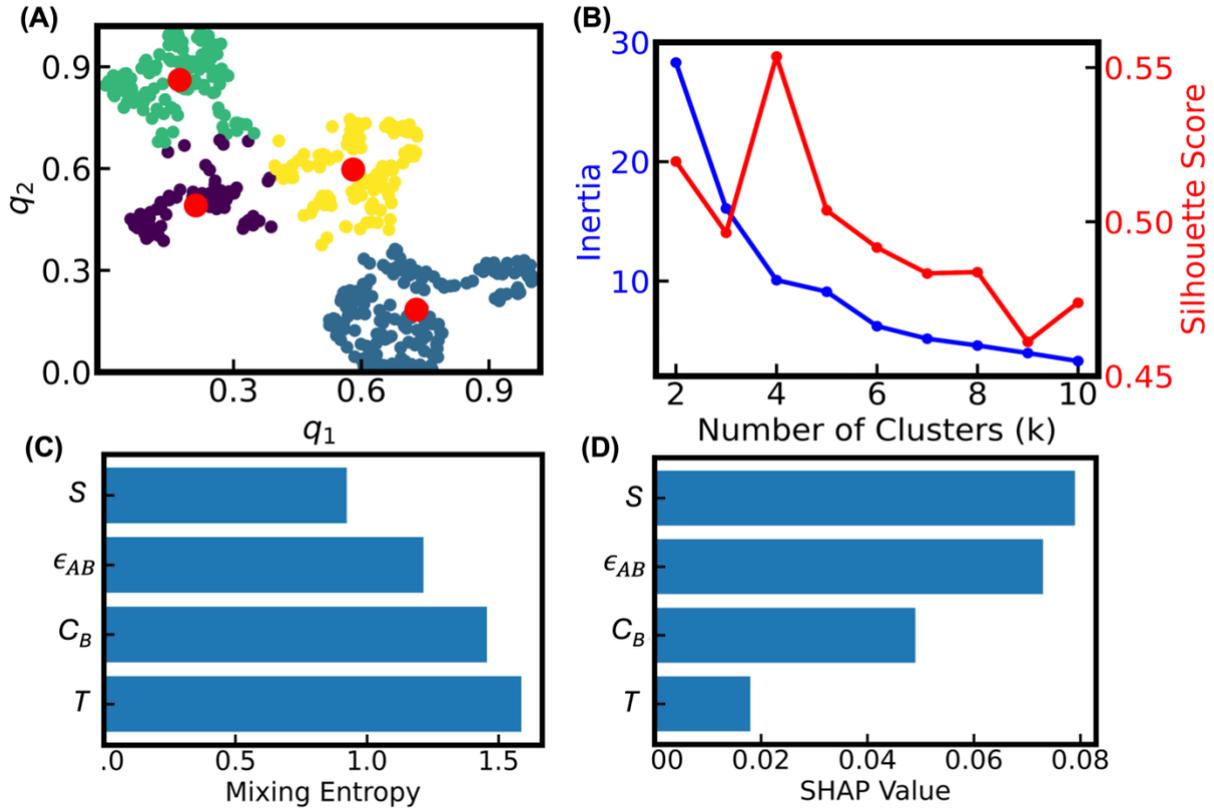

Figure 7: The UMAP of all the actual RDFs in a 2D representation is shown in (A). The inertia and Silhouette scores are plotted in (B) as a function of the data cluster size. The mixing entropy of all four features are ranked in (C). The SHAP analysis is shown in (D).

## V. Conclusions

AI offers exceptional promises to solve complex materials science problems and make rapid predictions of a material's behaviour, which are often resource intensive or intractable via physics-based methods. While AI has shown success in interpolative predictions, its reliability and limitations when applied beyond known data distributions are not yet well understood. Hence, a fundamental gap remains in understanding the scope of AI for predicting extrapolative regions in materials science, especially when these regions are governed by different physical principles. Addressing this challenge requires rigorous validation, uncertainty quantification, and domain-specific insights to ensure accurate predictions in unexplored material space. Also, it appears that the fidelity of an ML model decreases when the output dimension is very high. While AI has been effective in identifying patterns, predicting material properties, its ability to reveal underlying physical principles and mechanistic insights is still a subject of ongoing research. Here, we address these aspects of AI models in the context of predicting the RDFs of a binary mixture based on a relatively smaller number of training data points.

The low fidelity of the high dimensional regression task is tackled by discretizing an RDF into a large number of points. This approach reduces the output dimension to one as well as increase volume of training data. With discretized RDF data, we are able to build more accurate and efficient interpolative and extrapolative regression models. We use temperature, size ratio, concentration, cross-interaction as inputs for our AI model. Here, we perform extrapolation along the temperature axis. We carefully show model performance in predicting RDF in both lower and higher extrapolative temperature region. We systematically establish how model performance in the extrapolative region increases with increasing the size of training data. We show how the model's fidelity decreases when a phase transition occurs in the extrapolative region. Finally, we develop an UMAP to establish which input feature is most dominant in controlling the phase behavior and particle distribution in the binary LJ system. In addition, we perform the SHAP analysis to estimate the feature importance. Our analysis suggests that the particle size ratio is the most important feature of the RDFs of a binary fluid. Since binary LJ fluids can mimic a wide range of real materials, we expect this analysis to be useful in understanding and predicting properties of other fluid mixtures and material systems.

## Acknowledgements

This work is made possible by financial support from the SERB, DST, and Govt of India through a core research grant (CRG/2022/006926) and the National Supercomputing Mission's research grant (DST/NSM/R&D_HPC_Applications/2021/40). This research uses the resources of the Center for Nanoscience Materials, Argonne National Laboratory, which is a DOE Office of Science User Facility supported under the Contract DE-AC02-06CH11357.